\documentstyle[12pt,aasms4]{article}
\def\la{\mathrel{\hbox{\rlap{\hbox{\lower4pt\hbox{$\sim$}}}\hbox{$<$}}}}
\def\ga{\mathrel{\hbox{\rlap{\hbox{\lower4pt\hbox{$\sim$}}}\hbox{$>$}}}}

\received{}
\begin{document}

\title{X-ray Nova XTE J1550-564: Discovery of a QPO Near 185 Hz}
\author{Ronald A. Remillard}
\affil{Center for Space Research, Massachusetts Institute of Technology, Room 37-595, Cambridge MA 02139}
\authoremail{rr@space.mit.edu.edu}
\author{Jeffrey E. McClintock}
\affil{Harvard-Smithsonian Center for Astrophysics, 60 Garden St., Cambridge MA 02138}
\authoremail{jem@cfa.harvard.edu}
\author{Gregory J. Sobczak}
\affil{Astronomy Department, Harvard University, 60 Garden St., Cambridge MA 02138}
\authoremail{gsobczak@cfa.harvard.edu}
\author{Charles D. Bailyn}
\affil{Department of Astronomy, Yale University, P. O. Box 208101, New Haven, CT 06520}
\authoremail{bailyn@astro.yale.edu}
\author{Jerome A. Orosz}
\affil{Dept. of Astronomy and Astrophysics, The Pennsylvania State University, 525 Davey Laboratory, University Park, PA 16802}
\authoremail{orosz@astro.psu.edu}
\author{Edward H. Morgan}
\affil{Center for Space Research, Massachusetts Institute of Technology, Room 37-567, Cambridge MA 02139}
\authoremail{ehm@space.mit.edu}
\and
\author{Alan M. Levine}
\affil{Center for Space Research, Massachusetts Institute of Technology, Room 37-575, Cambridge MA 02139}
\authoremail{aml@space.mit.edu}

\begin{abstract}
   We have investigated the X-ray timing properties of XTE J1550-564
during 60 RXTE PCA observations made between 1998 September 18 and
November 28.  We detect quasi-periodic oscillations (QPOs) near 185 Hz
during four time intervals. The QPO widths (FWHM) are $\sim 50$ Hz,
and the rms amplitudes are $\sim 1$\% of the mean flux at 2-30 keV.
This is the third Galactic black hole candidate known to exhibit a
transient X-ray timing signature above 50 Hz, following the 67 Hz QPO
in GRS1915+105 and the 300 Hz QPO in GRO J1655-40. However, unlike the
previous cases, which appear to show stationary frequencies, the QPO
frequency in XTE J1550-564 must vary by at least $\sim 10$\% to be
consistent with observations. The occurrences and properties of the
QPO were insensitive to large changes in the X-ray intensity (1.5 to
6.8 Crab). However, the QPO appearance was accompanied by changes in
the energy spectrum, namely, an increase in the temperature and a
decrease in the normalization of the thermal component. The QPO is
also closely related to the hard X-ray power-law component of the
energy spectrum since the fractional amplitude of the QPO increases
with photon energy.  The fast QPOs in accreting black hole binaries
are thought to be effects of general relativity; however, the
relevance of the specific physical models that have been proposed
remains largely uncertain.  Low frequency QPOs in the range 3-13 Hz
were often observed.  Occasionally at high luminosity the rms QPO
amplitude was $\sim 15$\% of the flux, a level previously reached only
by GRS1915+105.  These extraordinary oscillations have a coherence
parameter ($\nu/\Delta\nu$) in the range 4-12 and are tied to the
power-law component in the energy spectrum.
\end{abstract}


\section{Introduction}

The X-ray nova and black hole candidate XTE J1550-564 was first
detected on 1998 September 6 (\cite{smi98}) with the All Sky Monitor
(ASM; \cite{lev96}) aboard the Rossi X-ray Timing Explorer (RXTE). It
is the brightest X-ray nova yet observed with RXTE. The ASM light
curve and further background information for this source are provided
in a companion paper by Sobczak et al. \nocite{sob99}(1999b; hereafter
paper I).  Extensive observations of the optical counterpart are
described by Jain et al. \nocite{raj99} (1999; hereafter paper
III). The 2.5-20 keV spectrum of XTE J1550-564 resembles the sources
that are dynamically established to be black hole binaries. The X-ray
and optical intensities of this source suggest a distance of roughly 6
kpc (paper I).

We present results from 60 RXTE observations of XTE J1550-564 that
were made between 12 and 83 days after the outburst began. Results
from earlier RXTE observations (outburst days 2--10) were reported by
Cui et al. (1999), who found that during the initial
rise (0.6--2.1 Crab) the source exhibited QPOs with a frequency that
systematically increased from 0.08 to 8 Hz as the X-ray flux
increased.  These QPOs were very strong, with rms amplitudes typically
$\sim$15\% of the mean flux over the full PCA band. In each
observation, the QPO amplitude (rms / mean flux) increased by a factor
of two between 2 and 30 keV. The temporal variability displayed by XTE
J1550-564 also resembles some of the black hole systems observed
during outburst (\cite{cui99}; paper I).

Herein we show a series of power spectra in which there are frequent
appearances of QPOs in the range 2-13 Hz. There are also a few
occasions in which we detect a high frequency QPO near 185 Hz which is
analogous to the stationary QPOs observed for two black hole
candidates: GRS1915+105 (67 Hz; Morgan, Remillard, \& Greiner 1997)
\nocite{mor97} and GRO J1655-40 (300 Hz; \cite{rem99b}).

\section{Observations and Analysis}

The times of the 60 RXTE pointed observations and a summary of some
X-ray properties of XTE J1550-564 are given in Table 1 of paper I.
We have analyzed the X-ray timing properties of XTE J1550-564 using
data from the PCA instrument (\cite{jah96}). Within the constraints of
spacecraft telemetry, we obtained moderately good time resolution in
at least a few energy bands by conducting the observations as
follows. In most cases, the PCA Event Analyzers (EAs) were configured to
deliver 122 $\mu$s time resolution in three broad energy bands, which
are approximately 2-6 keV, 6-12 keV, and 12-30 keV. The 30
keV boundary is an effective limit imposed by the source spectrum,
not by exclusion of high energy events in the data processing.
Lower time resolution was occasionally used to avoid possible telemetry
saturation due to high count rates: the time resolution was 250 $\mu$s
for observations \#4--6 (see paper I, Table 1) and 500 $\mu$s for
observations \#9--10.  In parallel, we usually used a fourth EA to
provide 8 energy bands with 4 ms time resolution within the energy
range 2-13 keV.

For each PCA observation, we computed power spectra for each of the 3
energy bands sampled with high time resolution and also for the 2--30
keV sum band. Power spectra were computed for every 256 s data
segment. Then for each observation and energy band, we averaged
together all of the 256 s power spectra.  We subtracted the
contribution from counting statistical noise, corrected for dead-time
effects as described by Morgan et al. (1997). \nocite{mor97} The
power spectra are normalized such that the power in each frequency bin
is the square of the rms amplitude divided by the mean count rate.  At
high frequencies, residual continuum power $\la 10^{-6}$ Hz$^{-1}$ is
likely to represent inaccuracies in our subtraction of statistical noise,
rather than source behavior.

We used a chi-squared minimization technique to derive the central
frequency and the width of an X-ray QPO. We fit each QPO feature with
a Lorentzian function, while the continuum on both sides of the QPO
was modeled with a power law function. On some occasions (e.g., see
Sept 21a,b below), it was necessary to add a quadratic term to the
relationship between log power density and log frequency, in order to
adequately model the curvature in the power continuum.  The QPO fit
parameters include the QPO central frequency ($\nu$) and the full
width at half maximum ($\Delta \nu$). The amplitude of the QPO,
expressed as a fraction of the mean count rate, is the square root of
the integrated power in the QPO feature. The central frequencies of
the 2--13 Hz QPOs are included in Table 1 of Paper I, while the
results for the fast ($\sim 185$ Hz) QPOs are given in Table 1 below.

\section{Results}

It can be seen in paper I that there are both short-term and long-term
variations in the intensity, X-ray spectrum, and QPO properties of XTE
J1550-564. Furthermore, the changes in timing and spectral parameters
are highly correlated (see Table 1 of paper I). To facilitate our
sensitivity to the high frequency QPOs, we average together power
spectra from sequential time intervals in which the changes in source
behavior are relatively minor. There are 12 such groups, and their
power spectra are shown in Figure~\ref{fig:pds1}.  The averaging has
smeared the low frequency QPOs (3--7 Hz) in panels d--g of
Figure~\ref{fig:pds1}, as can be discerned from Table 1 of paper I,
but this does not alter our conclusions below.

In our observations of 1998 September and October, XTE J1550-564 is
bright in X-rays with 2-30 keV intensity above 0.5 Crab. The majority
of the power spectra during this interval show a continuum that is
relatively flat below a few Hz; the power density values ($\sim$0.01)
imply $\sim$10\% rms variations at timescales longer than $\sim 0.2$
s. There is a sharp break in the continuum power near 5--10 Hz, with a
QPO feature near or somewhat above the break frequency. A second peak
is typically seen at the first harmonic frequency (2$\nu$), and a
weaker peak often appears at the first subharmonic (0.5$\nu$). These
power spectra resemble the earlier results for XTE J1550-564 that were
reported by Cui et al. (1999).

The QPOs in the range 2.6--13.1 Hz have the following characteristics.
While the source is bright, the detected QPOs have a coherence
parameter, $Q = \nu / \Delta\nu$, that is in the range 3.5 $< Q <
$12.0. However, as the source dims, there are infrequent detections of
broad QPOs with $Q \sim 1.6$ (see Figure~\ref{fig:pds1} and Table 1 of
paper I: Oct 29, Nov 9, and after Nov 20). The narrow QPOs are further
distinguished by their high amplitudes. In particular, the individual
observations between September 22 and October 13 generally yield rms
amplitudes of 8-14 \% of the mean count rate.  Thus the X-ray
luminosity, which is $\sim1.5 \times 10^{38} $($d$/6kpc)$^2$ erg
s$^{-1}$ (paper I), is modulated at 3--6 Hz with a crest-to-trough
ratio as high as 1.5. Previously, only the microquasar GRS1915+105 has
shown QPOs with such a large amplitude and a high luminosity
(\cite{mor97}).  These QPOs place significant constraints on physical
models designed to explain the power-law spectrum in accreting black
hole systems (e.g. \cite{mol96}; \cite{tit98}).  Further analyses of
these QPOs will be presented in a later publication.

While XTE J1550-564 is still in a bright state, the general shape of
the power spectrum diverges from the norm on September 19 and during
October 20--29. There is less power at low frequencies, and the
continuum can be very roughly described as a single power law with
index between 0.5 and 1.0.  More importantly, these observations
reveal an additional QPO near 185 Hz. This high frequency QPO is
strongest in panels b and h of Figure~\ref{fig:pds1}. These QPOs are
shown more clearly, along with the profile fits, in the left panels of
Figure~\ref{fig:fit185a}. The detections are significant at the level
of 6--7 $\sigma$, and the central frequencies are located at $184 \pm
6$ and $186 \pm 7$ Hz, respectively.

Power spectra and QPO fits in two different energy bands are shown for
the same two observations in Figure~\ref{fig:fit185b}, with fit
parameters given in Table~\ref{tab:qpo}. The uncertainty (1 $\sigma$)
in each parameter is calculated while fixing the other parameters at
their best-fit value. To investigate whether this method
underestimates the uncertainty, we plotted the surface contours of the
chi-square statistic for each pair of QPO parameters in the 2--30 keV
fits reported in Table 1. The asymmetries in these contours have only
minor significance, and they imply that the multi-parameter
uncertainties would be larger than the given ones by only 2--10\%.
For the QPO fits in the individual energy bands, where the statistics
are less reliable, the centroid frequency and width were fixed at the
values determined from the corresponding 2--30 keV fit. As shown in
Table~\ref{tab:qpo}, we are able to measure the QPO amplitude
independently at 2--6 keV and 6-12 keV, and there is clearly an
increase in the QPO amplitude with photon energy. In addition, there
is a weak indication that this trend continues into the 12--30 keV
band. These results for XTE J1550-564 are qualitatively consistent
with the increasing amplitude with energy seen in the 67 Hz QPO of GRS
1915+105 (\cite{mor97}). Thus, the fast X-ray oscillations in black
hole candidates are intimately tied to the hard X-ray component in the
energy spectrum.

Weaker high-frequency QPOs are seen during the days following each QPO
detection at 185 Hz. The QPO fits for September 21a,b and October
24-29 are shown in the right panels of Figure~\ref{fig:fit185a}, and
the fit parameters are included in Table~\ref{tab:qpo}. The detections
are significant at the level of 4--5 $\sigma$, and these QPOs are
centered at 161 $\pm$ 7 Hz and 238 $\pm$ 18 Hz, respectively. These
frequencies are inconsistent with 185 Hz at a confidence level $\sim
3\sigma$.  On the other hand the derived $Q$ values and the amplitudes
per energy band are consistent with the results for the stronger QPOs
at 185 Hz.  We must conclude that the fast QPO in XTE J1550-564 shows
significant variations in frequency, the first such evidence among the
three black hole candidates that display high frequency QPOs. At the
90\% confidence level, the high frequency QPO in XTE J1550-564 must
vary by $\sim \pm 10$ \% to be consistent with these observations. We
further note that while this paper was under review and as the X-ray
outburst of XTE J1550-564 continued, there were reports of even larger
variations in frequency, as a QPO appeared at 284 Hz and then settled
back to 182 Hz (\cite{hom99}; \cite{rem99a}).

\section{Discussion}

Spectral analyses of these 60 RXTE observations (paper I) were made
using the standard model composed of a disk blackbody plus a power-law
component.  The results characterize the spectral evolution of XTE
J1550-564 through the peak and initial decay phases of the
outburst. The September 19 detection of the QPO at 184 Hz is
coincident with a huge (6.8 Crab) X-ray flare that lasted between one
and two days. On the other hand, the October 20-23 detection occurs
during a very minor increase in intensity ($\sim$1.5 Crab; note the
small arrows in Fig. 1 of paper I). Thus intensity is a poor
diagnostic of the conditions that produce the fast QPO. Far more
useful indicators are the color temperature ($T_{col}$) and
normalization of the disk component, which is proportional to the
square of the color radius ($R_{col}$). The data in columns 8 \& 9 of
Table 1 in paper I show that all of the fast QPOs occur when the color
temperature is relatively high ($T_{col} \ga 0.84$ keV), while the
disk color radius (scaled to 6 kpc with a pole-on view) is relatively
small ($< 40$ km).  We further note that the power law component
contributes more than half of the total X-ray luminosity during all of
our observations that occur before October 29. This entire scenario,
i.e. the detection of fast QPOs when the inner disk appears small and
hot while the hard X-ray power law is very strong, is the same suite
of conditions that accompanied the 300 Hz QPO in GRO J1655-40
(\cite{rem99b}; \cite{sob99a}). Furthermore, the rms amplitude of the
fast QPO in XTE J1550-564 ($\sim$ 1\%) and its broad profile ($Q \sim
3.5$) are also very similar to the values derived for the 300 Hz QPO
seen in GRO J1655-40.

As noted in the previous studies of fast QPOs in black hole
candidates, it is natural to hypothesize that these msec timing
signatures in the emission from very hot material represent a
fundamental timescale of the inner disk. However the cause of this QPO
appears to involve both the disk and power-law components, since the
onset of the QPO is related to the temperature of the disk, while the
energy dependence of the QPO amplitude implies that the oscillation is
tied to the power-law component.

Two different physical models have been advanced for these
high-frequency QPOs, and both are effects of general relativity that
depend on the mass and spin of the black hole.  In the case of
``Lense-Thirring'' precession, or the ``frame-dragging'' model,
vertical structure in the inner disk gives rise to a relativistic
precession, and the precession frequency could impose a timing
signature on the X-ray emission (\cite {ste98}; \cite{cui98}; \cite{mer98}).
Merloni et al. (1998) have shown that fast precession
($\nu > 10$ Hz) signifies a rapidly rotating black hole with spin
parameter $a > 0.5$. A change in the precession frequency might
correspond to a shift in the radius of peak X-ray emissivity.
Computations by Markovic \& Lamb (1998) \nocite{mar98} indicate that
some high-frequency modes of this oscillation may survive against
strong damping. However, the means to initially excite these modes is
unclear. Furthermore, it is unclear how precession of the inner disk
would produce a QPO amplitude that increases with photon energy (Table
1).

An alternative model is the ``diskoseismic'' oscillation in which
normal mode oscillations are trapped via relativistic effects in the
inner disk (\cite{per97}; \cite{wag98}; see also \cite{che95}).  This
model predicts oscillations in density and disk thickness that could
produce observable effects in the X-ray emission. The oscillation
frequency depends on the mass and spin of the black hole, as well as
the radius of peak emissivity. Some of the oscillation modes also
depend on the thickness of the disk, which is expected to depend on
the mass accretion rate (\cite{wag98}). Therefore this model can also
accomodate observed changes in the QPO frequency.  Again, the
mechanism by which these oscillations would reproduce the energy
dependence of the QPO amplitude is not evident.

Clearly, the accumulation of numerous high-quality measurements of 
fast QPOs from a variety of black hole systems is a necessary step in
developing a sound physical theory for this phenomenon. Guidance on
interpreting the fast QPO of XTE J1550-564 may come from optical
observations which may yield a determination of the mass of the black
hole (paper III).

\acknowledgments 
This work was supported, in part, by NASA contract NAS5-30612 and 
NASA grant NAG5-3680. Partial support for J.E.M. was provided by the 
Smithsonian Institution Scholarly Studies Program.

\clearpage

\begin{center}
{\bf Figure Legends}
\end{center}

\figcaption[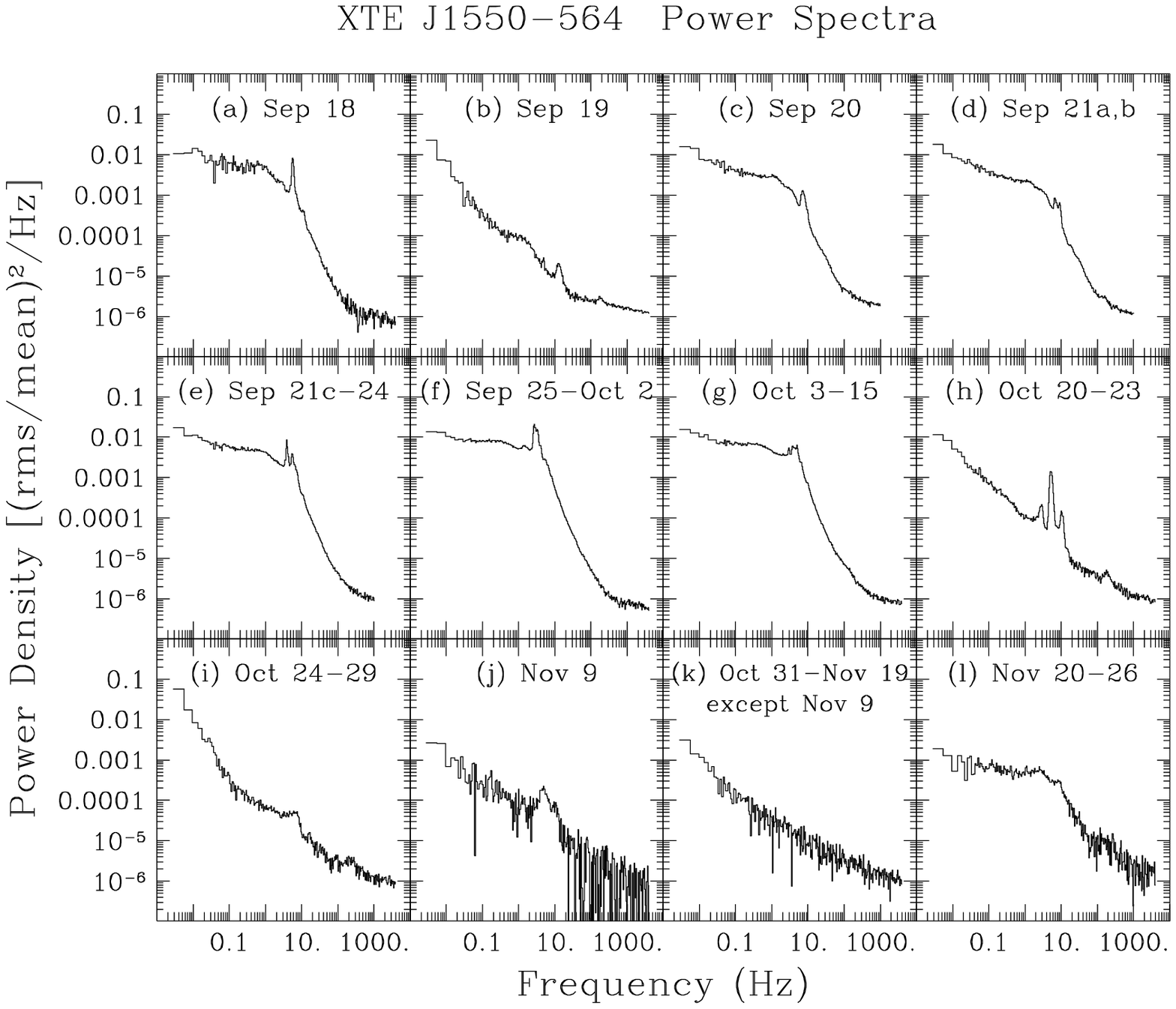]{PCA power spectra of XTE J1550-564 in 12 
sequential observation groups. The power due to counting statistics, 
corrected for instrument dead time, has been subtracted. \label{fig:pds1}}

\figcaption[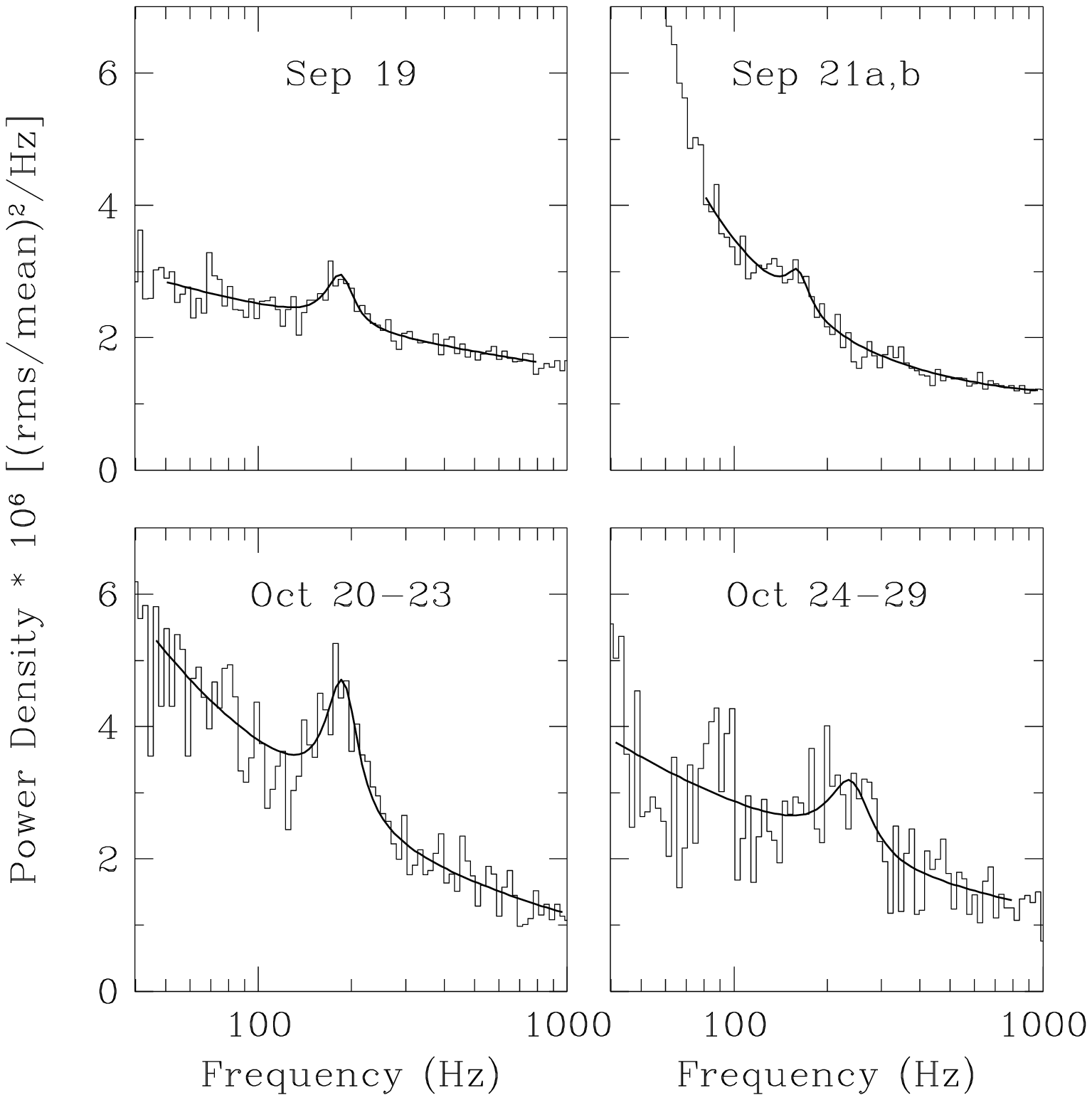]{An expanded region of the PCA power spectra 
(2--30 keV) for the observations in which there are significant QPO 
detections at high frequency. Here the power density is plotted 
in linear units. The fits to the QPO feature and the power continuum 
are shown with a smooth line.
\label{fig:fit185a}}

\figcaption[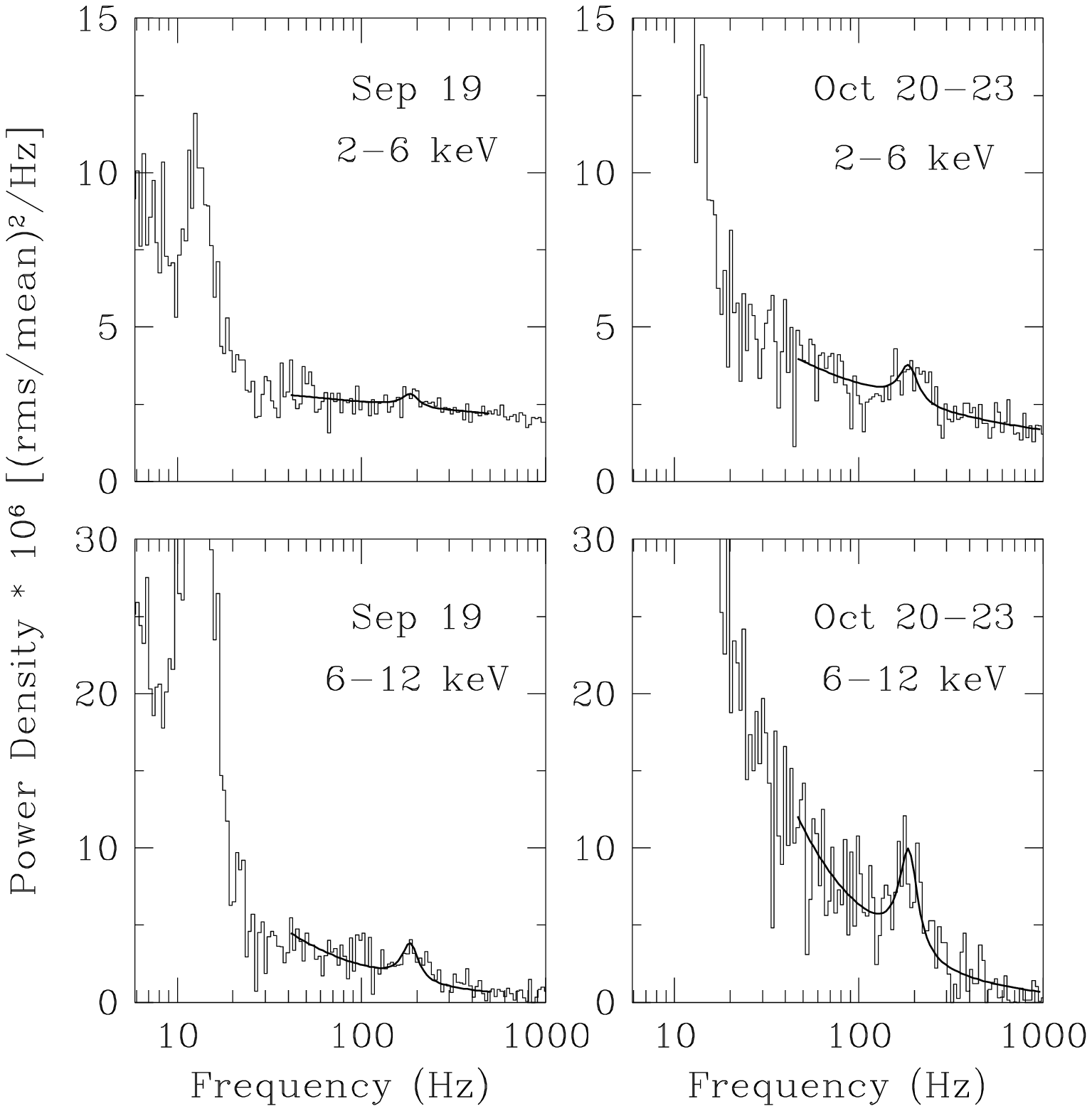]{The power spectra and QPO fits are shown in
two energy channels for the strongest QPO detections near 185 Hz. 
The central frequency and QPO width are fixed at the values 
determined for the sum band (2--30 keV). In each observation, 
the amplitude of the 185 Hz QPO increases with photon energy,
as does the 13 Hz QPO seen on September 19. \label{fig:fit185b}}

% ---------------------------------------------------------------
% -------------------     TABLES                 ----------------
% ---------------------------------------------------------------
 
\clearpage
 
\begin{deluxetable}{cccccc} \footnotesize \tablecaption{High Frequency QPOs in XTE J1550-564 \label{tab:qpo}} \tablewidth{0pt}
\tablehead{ \colhead{Date} & \colhead{range} & \colhead{$\nu$} & \colhead{$\Delta\nu$} & \colhead{QPO}  & \colhead{$\chi_\nu^2$}  \\ 
\colhead{(mm dd)} & \colhead{(keV)} & \colhead{(Hz)} & \colhead{(Hz)} & \colhead{amplitude} & \colhead{} }
\startdata 
09 18 &  2-30 & 183.9 (6.0) & 46.7 (9.3) & 0.0068 (0.0010) & 0.91 \\
09 18 &  2-6  & 183.9 (0.0) & 46.7 (0.0) & 0.0051 (0.0016) & 0.82 \\
09 18 &  6-12 & 183.9 (0.0) & 46.7 (0.0) & 0.0122 (0.0026) & 1.00 \\
09 18 & 12-30 & 183.9 (0.0) & 46.7 (0.0) & 0.0178 (0.0065) & 1.15 \\
\\
09 21a,b &  2-30 & 161.1 (6.5) & 35.1 (11.5) & 0.0052 (0.0012) & 1.07 \\
09 21a\tablenotemark{a}
   &  2-6  & 161.1 (0.0) & 35.1 (0.0)  & 0.0039 (0.0022) & 0.80 \\
09 21a   &  6-12 & 161.1 (0.0) & 35.1 (0.0)  & 0.0106 (0.0024) & 1.22 \\
09 21a   & 12-30 & 161.1 (0.0) & 35.1 (0.0)  & 0.0058 (0.0130) & 0.82 \\
\\
10 20-23 &  2-30 & 186.2 (7.0) & 53.7 (13.0) & 0.0120 (0.0020) & 1.11 \\ 
10 20-23 &  2-6  & 186.2 (0.0) & 53.7 (0.0)  & 0.0088 (0.0023) & 1.07 \\ 
10 20-23 &  6-12 & 186.2 (0.0) & 53.7 (0.0)  & 0.0206 (0.0042) & 1.25 \\ 
10 20-23 & 12-30 & 186.2 (0.0) & 53.7 (0.0)  & 0.0253 (0.0090) & 0.94 \\ 
\\
10 24-29 &  2-30 & 237.6 (17.9) & 95.5 (26.5) & 0.0120 (0.0026) & 1.06 \\ 
10 24-29 &  2-6  & 237.6 (0.0)  & 95.5 (0.0)  & 0.0097 (0.0030) & 0.86 \\ 
10 24-29 &  6-12 & 237.6 (0.0)  & 95.5 (0.0)  & 0.0272 (0.0060) & 0.71 \\ 
10 24-29 & 12-30 & 237.6 (0.0)  & 95.5 (0.0)  & 0.0510 (0.0296) & 0.94 \\ 

\enddata
\tablenotetext{a}{The observation on September 21b yielded only the sum band at high time resolution; therefore, the analysis of the QPO amplitude in individual energy bands is limited to results from September 21a only.}
\end{deluxetable}

\clearpage
\plotone{rrpds1.ps}
\clearpage
\plotone{rrpds2.ps}
\clearpage
\plotone{rrpds3.ps}

\end{document}